\pdfoutput=1
\documentclass{emulateapj}

\usepackage{natbib}
\usepackage{epstopdf}

\shorttitle{The IR spectra of dPAHs}
\shortauthors{Mackie et al.}

\begin{document}


\title{Characterizing the infrared spectra of small, neutral, fully
 dehydrogenated PAHs}


\author{C.~J.~Mackie\altaffilmark{1,2}, 
  E.~Peeters\altaffilmark{1,3},
  C.~W.~Bauschlicher,Jr\altaffilmark{4}, 
  and
  J.~Cami\altaffilmark{1,3}}

\altaffiltext{1}{Department of Physics and Astronomy, University of
 Western Ontario, London, ON N6A 3K7, Canada}\email{Electronic address: mackie@strw.leidenuniv.nl}
\altaffiltext{2}{Leiden Observatory, Leiden University, PO Box 9513,
 NL-2300RA, Leiden, The Netherlands }
\altaffiltext{3}{SETI Institute, 189 Bernardo Ave, Mountain View, CA 94043, United States}
\altaffiltext{4}{NASA Ames Research Center, MS 245-6, Moffett Field,
 CA 94035, United States}
\slugcomment{Accepted for publication in \apj}


\begin{abstract}
We present the results of a computational study to investigate the
infrared spectroscopic properties of a large number of polycyclic
aromatic hydrocarbon (PAH) molecules and their fully dehydrogenated
counterparts. We constructed a database of fully optimized geometries
for PAHs that is complete for eight or fewer fused benzene rings, thus
containing 1550 PAHs and 805 fully dehydrogenated aromatics. A large
fraction of the species in our database have clearly non-planar or
curved geometries. For each species, we determined the frequencies and
intensities of their normal modes using density functional theory
calculations. Whereas most PAH spectra are fairly similar, the spectra
of fully dehydrogenated aromatics are much more diverse. Nevertheless,
these fully dehydrogenated species show characteristic emission features
at 5.2$\mu$m, 5.5$\mu$m and 10.6$\mu$m; at longer wavelengths, there
is a forest of emission features in the 16--30$\mu$m range that
appears as a structured continuum, but with a clear peak centered
around 19$\mu$m. We searched for these features in Spitzer-IRS spectra
of various positions in the reflection nebula \object{NGC 7023}. We
find a weak emission feature at 10.68$\mu$m in all positions except
that closest to the central star. We also find evidence for a weak
19$\mu$m feature at all positions that is not likely due to
C$_{60}$. We interpret these features as tentative evidence for the
presence of a small population of fully dehydrogenated PAHs, and
discuss our results in the framework of PAH photolysis and the
formation of fullerenes. 
\end{abstract}




\section{Introduction}

The infrared (IR) spectra of almost all astronomical objects, ranging
from young stellar objects to planetary nebulae, and from galactic
nuclei to the general ISM of galaxies, are dominated by prominent
emission bands at 3.3, 6.2, 7.7, 8.6, 11.2, 12.7, and 16.4 $\mu$m, as
well as weaker features at 3.4, 3.5, 5.25, 5.75, 6.0, 6.9, 7.5, 10.5,
11.0, 13.5, 14.2, 17.4, and 18.9$\mu$m. It has long been suggested
\citep[e.g.][]{1984A&A...137L...5L, 1985ApJ...290L..25A} that these
bands are caused by the combined emission of a population of polycyclic
aromatic hydrocarbon molecules (PAHs) with sizes of typically 50 C
atoms. It is now generally accepted that this is indeed the case, and
that PAHs lock up some 10--20\% of the cosmic carbon \citep[see
 e.g.][]{2008ARA&A..46..289T, 2014IAUS..297..187P}\footnote{Note that
 over time, it has become clear that the astronomical emission does
 not correspond to PAHs as defined by the International Union of Pure and Applied Chemistry -- i.e. only C and H --
 but rather to polycyclic aromatic compounds that can include
 hetero-atoms and substituted species or groups, correspond to
 various degrees of (de-)hydrogenation etcetera.}. Absorption of a
single UV photon through an electronic transition, followed by internal energy
redistribution, leaves PAHs in highly excited vibrational states. 
They then cool by fluorescence, emitting
IR photons at the frequencies corresponding to these vibrational modes
\citep{ATB89, Bakes_and_Tielens:1994, 2001ApJ...556..501B,
 2001ApJ...560..261B}. Since some vibrational modes (in particular
stretching or bending of C-C and C-H bonds) are common to most PAHs
and also occur at similar frequencies, a population of PAHs will emit
very strongly at those common frequencies. As a consequence, the
observed emission bands are overall fairly similar in appearance in
different lines of sight, and these IR PAH bands (from now on simply referred to as the IR bands)
can thus not be used to identify individual PAH species.

However, there are clear (albeit sometimes subtle) variations in the
precise peak positions, line profiles and band intensity ratios of the
IR bands in different astronomical sources or even spatially within
extended sources. From a comparison to laboratory spectra and
theoretical calculations, these variations can be understood as
changes in the properties of the underlying PAH population such as
charge state, precise chemical make-up or size \citep{peeters:prof,
 Hony:oops:01, vandiedenhoven, 2005ApJ...632..316H,
 Cami:PAHfit:Toulouse}. 

A full understanding of these variations and the further development
of quantitative diagnostic tools depends on the availability of
spectral properties for large subsets of PAH species. Experimentally
obtained and theoretically calculated spectra of hundreds of PAH
species have become more available in recent years \citep[see e.g.~the
 NASA Ames PAH database; ][]{2010ApJS..189..341B, PAHdbv2}. Model
fits using such collections of PAH spectral properties with calculated
fluorescence spectra can reproduce the observed emission and their
variations remarkably well
\citep[e.g.][]{1999ApJ...511L.115A,Cami:PAHfit:Toulouse,PAHdbv2}. However,
these databases are still highly biased and incomplete, and it is not
immediately clear whether all astrophysically relevant subclasses are
included. 

\bigskip

PAHs share many properties with another class of large aromatic
species that was only recently detected in
space. \citet{Cami:C60-Science} detected several emission bands (most
notably at 7.0, 8.5, 17.4 and 18.9$\mu$m) in the infrared spectrum of
the peculiar planetary nebula Tc~1 and identified them with the
vibrational modes of the fullerene species C$_{60}$ and C$_{70}$. The
characteristic C$_{60}$ bands (especially the 17.4 and 18.9$\mu$m
bands) have since been found in many objects, including several
evolved stars \citep[in our Milky Way galaxy as well as in the
 Magellanic Clouds; see e.g.][]{ Garcia-Hernandez:PN,
 Garcia-Hernandez:MC, Garcia-Hernandez:RcrB, Gielen:C60p-AGB,
 ZhangKwok:proto-PNC60,2013ApJ...764...77O,Masaaki:C60ONe} but also
interstellar environments
\citep{2010ApJ...722L..54S,2011MNRAS.410.1320R, 2012ApJ...747...44P,
 2012ApJ...753..168B} and young stellar objects
\citep[YSOs,][]{Roberts:C60}. Of particular interest here is the
spectrum of the Reflection Nebula (RN)
NGC~7023. \citet{2007ApJ...659.1338S} first suggested C$_{60}$ as a
possible carrier for the band at 18.9$\mu$m in this source; they later
searched and detected also the 7.0$\mu$m band
\citep{2010ApJ...722L..54S}. Furthermore, near the central star,
\citet{2013A&A...550L...4B} found several weak bands that are
coincident with the vibrational modes of C$_{60}^+$. 

One important aspect about cosmic fullerenes that is not understood
yet is their precise formation route. Laboratory experiments have
shown that fullerenes can self-assemble from a hot carbon vapour in a
bottom-up process \citep{1985Natur.318..162K, 2009ApJ...696..706J,
 Dunk:CNG, Dunk:metallofullerenes}. However, densities and
temperatures in the circumstellar environments where fullerenes have
been detected are generally too low for such a bottom-up route to
complete in reasonable timescales. But other routes may exist as
well. \citet{ISI:000277926800011} showed that a graphene sheet rolls
up and forms C$_{60}$ molecules after losing an edge carbon
atom. Based on these results, \citet{2012PNAS..109..401B} suggested
that in space, similar processes may convert large PAHs into
fullerenes. They argue that UV irradiation first results in fully
dehydrogenated PAHs (which we refer to here as dPAHs\footnote{Even
 though fully dehydrogenated PAHs are technically speaking no longer
 considered a hydrocarbon, we still refer to these neutral
 dehydrogenated PAHs as dPAHs since they originate from the act of
 dehydrogenating PAHs in our exercise. Alternatives could be PAdHs,
 PACs or graphene-like molecules. }) and subsequently causes carbon
loss which results in the formation of a pentagonal ring that induces
curving up of the structure. Through isomerization the dPAH then
rearranges to finally arrive at the very stable fullerene
configuration on very short time-scales. While such a formation route
could be at play in RNe such as NGC~7023 where large PAHs could be
abundantly available as starting materials, it is probably not the
dominant route in PNe where we see fullerenes exclusively in the
low-excitation objects \citep[][]{Masaaki:C60ONe,
 Cami:DIBconf,2014ApJ...791...28S}. In those environments, fullerenes
could maybe form through aggregation and closed network growth
\citep{Dunk:CNG} from carbon clusters such as small dPAHs
\citep{Cami:DIBconf}. Alternative photoprocessing routes starting from
Hydrogenated Amorphous Carbon (HAC) grains have been proposed as well
\citep{Garcia-Hernandez:PN,Elisabetta:arophatics}.

In either of the two formation routes starting from PAHs, dPAHs are a
clear intermediate step. Detecting the clear signatures of these dPAHs
could thus further elucidate some of the pathways toward the
processing and evolution of interstellar PAHs and the possible
formation of fullerenes. However, as of yet, only a handful of
individual dPAH species have been studied in great detail \citep[see
 e.g][]{charlie:dehydrogenated}, and it is thus not clear what (if
any) could be a clear detection signature for the family of dPAHs as a
whole. Consequently, it is not clear whether dPAHs are a component of
the interstellar aromatic molecular species.
\bigskip

Here, we aim to characterize the changes in the infrared spectra of
PAHs as they become fully dehydrogenated. To do this, we produced a
comprehensive database of the structure and vibrational modes of PAH
and dPAH species containing eight or fewer benzenoid rings. We used
this database to study the spectral differences between PAHs and
dPAHs, and to describe characteristic spectral features that could be
indicative of the presence of dPAHs. Finally, we studied the
Spitzer-IRS spectra of NGC~7023 at several positions between the
PAH-dominated areas and the fullerene-rich zone to evaluate whether
dPAHs could be present. 

We present in \S~\ref{Sec:Methods} the computational methods used to
build the database. We present an overview of the resulting PAH and
dPAH molecular structures in \S~\ref{Sect:Structures} and the spectral
analysis and characterization in
\S~\ref{Sect:spectral_analysis}. Finally, the comparison between our
calculated spectra and observations of NGC~7023 is presented in
\S~\ref{Sec:NGC7023}.


\section{Computational Methods}
\label{Sec:Methods}

The goal of this work is to produce an unbiased database of the IR
spectra of small PAHs and their dehydrogenated counterparts in order
to characterize the changes to the spectra upon dehydrogenation. It
was for this reason that we decided to produce a complete database of
ever single possible PAH and corresponding dPAH species up to a given
number of benzenoids. Other papers have looked at the effect on an IR
spectrum of a PAH after removing all hydrogens, such as in
\citet{charlie:dehydrogenated}, but these only looked at a handful of
species at a time. By studying a large and complete database of
species, we can make better predictions of the dPAH family as a whole,
and especially so for the subset of smaller dPAHs. 

\subsection{Enumerating PAH geometries}

We used a recursive method to determine all possible PAH geometries.
To produce all PAHs/dPAHs with $h$ benzenoid members we add an
additional benzenoid ring to each open edge of all species with $h-1$
benzenoids. Starting with a benzene seed we iterate this process up to
species where $h$ equals eight. To simplify the procedure each benzene
ring is treated as a hexagon so that the problem becomes producing all
possible simple polyhexes. The connectivity between each open face of
each hexagon is then easily mapped by arranging them in a grid. After
all $h$ membered polyhexes are found a procedure then converts the
vertexes of each hexagon to the coordinates representing the position
of carbon atoms, adopting a C-C bond length at this point of 1.4\AA.

This method however produces duplicate species. To remove these
duplicates each of the carbon skeletons needs to be compared to the
others with the same number of C atoms. We therefore calculated the
moment of inertia tensor of each carbon skeleton, and used the
resulting eigenvectors to produce a rotation matrix that rotates the
species to lie along its principal axes. This causes each duplicate
species to be rotated into the same orientation after which each
coordinate of each carbon atom can be compared. It should be noted
that reflections across axes also need to be compared as this method
of rotating to the principal axes is ambiguous when it comes to
direction of positive and negative directions. Moreover, if the moment
of inertia is equal between two axes, we have to take axis
interchangeability into account. After all duplicates are removed this
method can be repeated using the outputted unique polyhexes as input
for the next iteration.

\begin{figure}[t]
\resizebox{\hsize}{!}{\includegraphics{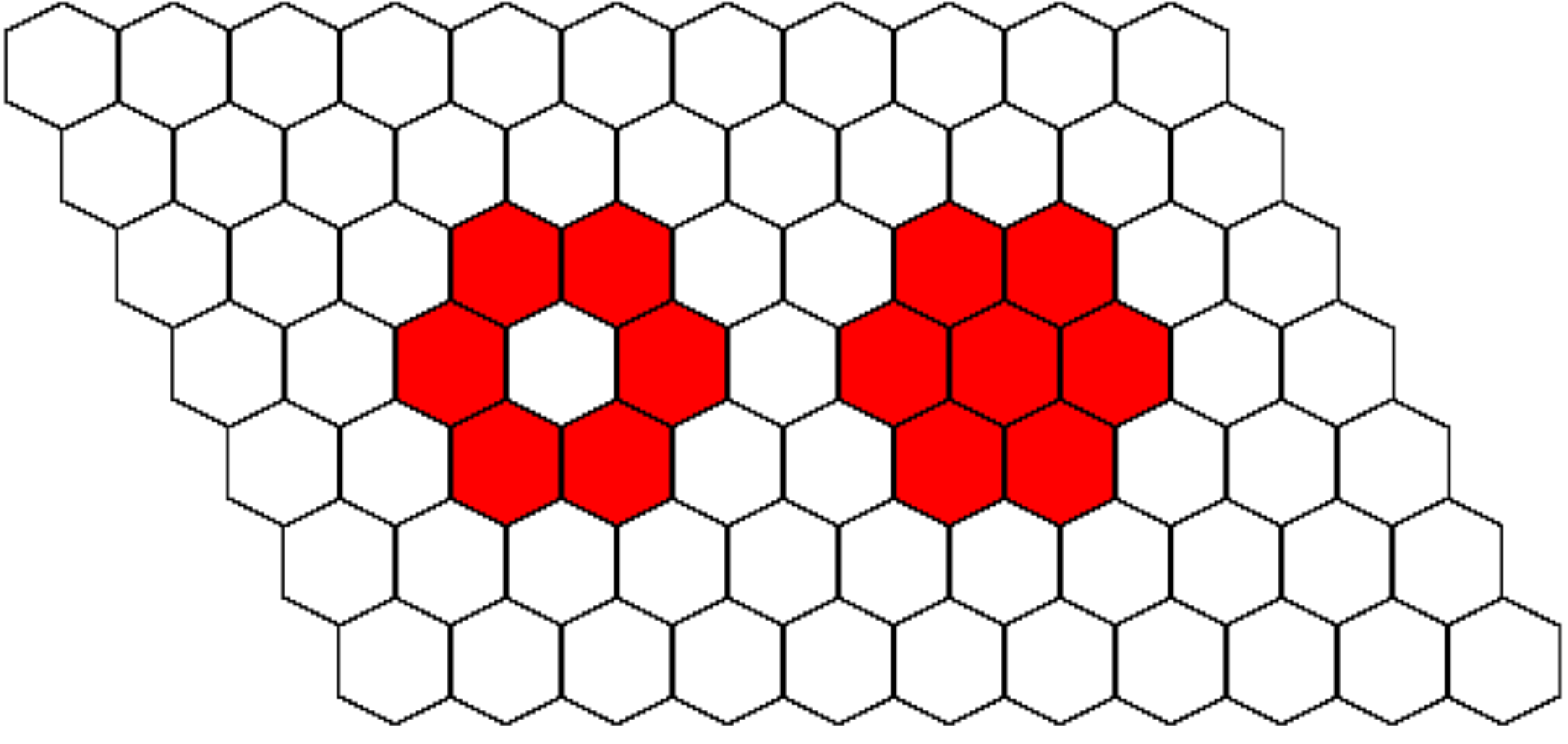}}
\caption{\label{Fig:duplicates_illustration}An illustration of our
 method to enumerate hexagons, and of the duplicates that arise when
 considering structures with a different number of hexagons. Shown in
 red on the left is a structure obtained by using 6 hexagons; on the
 right is a structure obtained using 7 hexagons. When considering
 only the vertices (i.e. the location of the carbon atoms), both
 structures are of course the same - the structure of coronene. Such
 issues arise starting from 5 benzenoid rings. }
\end{figure}

\begin{table}[t]
\centering
\caption{Number of PAH species with a number of benzenoid rings equal to $h$}
\label{table:nbenz}
\begin{tabular}{ccc}
\hline
$h$ & Dualist graph polyhexes & PAHs this paper\\
\hline
2 & 1 & 1 \\
3 & 3 & 3\\
4 & 7 & 7\\
5 & 22 & 21 \\
6 & 81 & 75 \\
7 & 331 & 286 \\
8 & 1436 & 1157 \\
9 & 6510 & 4886 \\
10 & 30129 & 20890\\
\hline
\end{tabular}
\end{table}

Some ambiguity occurs though when $h$ is equal to or greater than
five, and is best illustrated by the situation depicted in
Fig.~\ref{Fig:duplicates_illustration}. The two structures depicted in
that figure are obtained for $h=6$ and $h=7$ respectively. However,
since we are concerned with the hexagon {\em vertices} here (which
correspond to the positions of the C atoms) and not the hexagon faces,
both structures are identical for our purposes, and the $h=6$
structure really corresponds to 7 benzenoid rings. More generally, for
$h\ge5$, our method produces species out of $h$ rings that we actually
consider to be $h+1$ membered rings; for $h\ge7$ we get species with
$h+2$ membered rings etcetera. Once all initial polyhexes are made the
program then goes back and compares all species for uniqueness once
again to check for duplicates across differing $h$ numbers. For the
sake of classifying these structures, the final number of benzenoid
rings needs to be recounted. We assigned each duplicate species to the
class corresponding to the highest number of benzenoids counted --
e.g. $h=7$ for the example of coronene shown above.

We calculated all possible unique PAH structures for $h\le10$ (see
Table~\ref{table:nbenz}). It should be noted that the problem of
enumerating all possible arrangements of a given number of adjoining
hexagons has been investigated in mathematics using a method called
the {\em dualist tree} \citep[see
 e.g.][]{Nikolic:dualist_tree}. However, the dualist tree method
counts the {\em faces} of hexagons and not the vertices. Thus, what we
consider to be duplicate PAH across differing $h$ iterations are
actually considered unique in the dualist tree method (see e.g. the
coronene example in Fig.~\ref{Fig:duplicates_illustration}). Therefore
the dualist tree method only sets an upper bound on the number of
possible PAH structures for a given number of benzenoid rings, and
thus the number of possible structures found in this paper are lower
than the numbers predicted by the dualist tree method
\citep{Nikolic:dualist_tree}.

The polyhexes obtained in this way correspond to the initial
geometries for the dPAHs. However, we will also need the geometries
for the regular hydrogenated PAHs, i.e. we need to also determine the
position of the H atoms. For our starting geometries, we used a simple
computational technique that accurately places each hydrogen.
Starting with the center of each hexagon the hydrogen atoms are placed
radially out past each carbon atom. However, if a hydrogen atom ends
up too close to a carbon atom as a result of this operation (such as
in the body of a PAH), it is removed from the structure. This then
results in the initial geometries for the regular PAHs.

\subsection{Calculating spectral properties}

\begin{table}[t]
\centering
\caption{Ground state spin state of PAHs and dPAHs}
\label{table:spin}
\begin{tabular}{ccc}
\hline
Spin state & \% of total PAHs & \% of total dPAHs \\
\hline
Singlet & 54 & 27 \\
Doublet & 43 & 0 \\
Triplet & 3 & 37 \\
Quadruplet & 0 & 0 \\
Quintet & 0 & 28 \\
Sextet & 0 & 0 \\
Septet & 0 & 7 \\
Octet & 0 & 0 \\
Nonet & 0 & 1 \\
\hline
\end{tabular}
\end{table}

We then computed the spectral properties of the PAHs and dPAHs using
density functional theory (DFT). For all species, we used the Becke
three-parameter, Lee-Yang-Parr (B3LYP) exchange-correlation functional \citep{1993JChPh..98.5648B,lyp, stephens_gaussian} together
with the 4-31G basis set \citep{1984JChPh..80.3265F}. The reliability of the B3LYP/4-31G approach has been considered several times previously in regards to PAHs. The average absolute
and maximum errors in the scaled B3LYP/4-31G frequencies of naphthalene compared with experiment are 4.2 and 12.5 cm$^{-1}$,
respectively; these are to be compared with 2.8 and 8.1~cm$^{-1}$ for the very large cc-pVQZ basis set \citep{BauschlicherRica:2010}. If the generalized
gradient approximation is used instead of the hybrid B3LYP functional, the errors increase. A comparison of the
computed and experimental frequencies for 13 PAHS ranging in size from C$_{22}$H$_{12}$ to C$_{50}$H$_{22}$ again supported the use of the
B3LYP/4-31G approach \citep{Boersma:2010}. The scaled B3LYP/4-31G, B3LYP/6-31G*, BP86/4-31G, and BP86/6-31G harmonic
frequencies for C$_{60}$ are in good mutual agreement and
in good agreement with the 4 IR and 10 Raman bands from experiment \citep{Bauschlicher:sub}; the average absolute error
for these four levels are 7.8, 4.2, 10.0, and 8.5 cm$^{-1}$, respectively, while the maximum errors are 18.3, 12.9, 27.1,
and 21.6 cm$^{-1}$.
This approach was also used for the majority of spectra in the NASA Ames PAH
database \citet{2010ApJS..189..341B, PAHdbv2}, and yields results that
are accurate enough for our purposes, while not being too
computationally expensive. All calculations were carried out using the
software suite Gaussian~09 \citep{GAUSSIAN..09}.

We first formatted the atom coordinates of each PAH and dPAH for input
into Gaussian~09. For each species, we then fully optimized the
geometry, and subsequently calculated the harmonic vibrational
frequencies. It has been shown that spin state can have an affect on
line positions and intensities attributed to solo site electron pair
interactions, and that the lowest spin state for dPAHs is not
necessarily the ground state \citep{charlie:dehydrogenated}.
Therefore, we also explored higher spin states for each PAH and dPAH
molecule; due to the large number of species in our database however,
we could only carry out the most basic test. Each species thus had
its spin state increased; if this lowered the total energy (including
the zero-point energy), we increased the spin state again until the
spin state was found with the lowest energy. This was than taken as
the ground state configuration. If a transition state was found it
was disregarded, even though it is possible that geometry changes with
higher spin states could result in lower energy. This had general
affect of weakening the intensities of the bands between 5 and 15$\mu$m
by half relative to the intensities of the bands between 15 and
30$\mu$m. No significant shifts in peak positions were observed.
Table \ref{table:spin} summarizes the ground state spin states found
for the PAHs and dPAHs. Note that neutral dPAHs always have an even
number of electrons.

Although we calculated the initial structures for $h\le10$, we only
calculated the spectra for all PAHs and dPAHs for $h\le8$ due to
computational limitations. One aspect that could not be fully
automated was the process of producing optimized geometries, since any
deviations from planarity (e.g. the repulsion due to two H atoms in
close proximity) needed to be manually addressed. Moreover, when these
deformations occur in multiple places on one particular species, an
intuitive call about the stability of a particular stereochemistry has
to be made.

\begin{figure}[t]
\resizebox{\hsize}{!}{\includegraphics{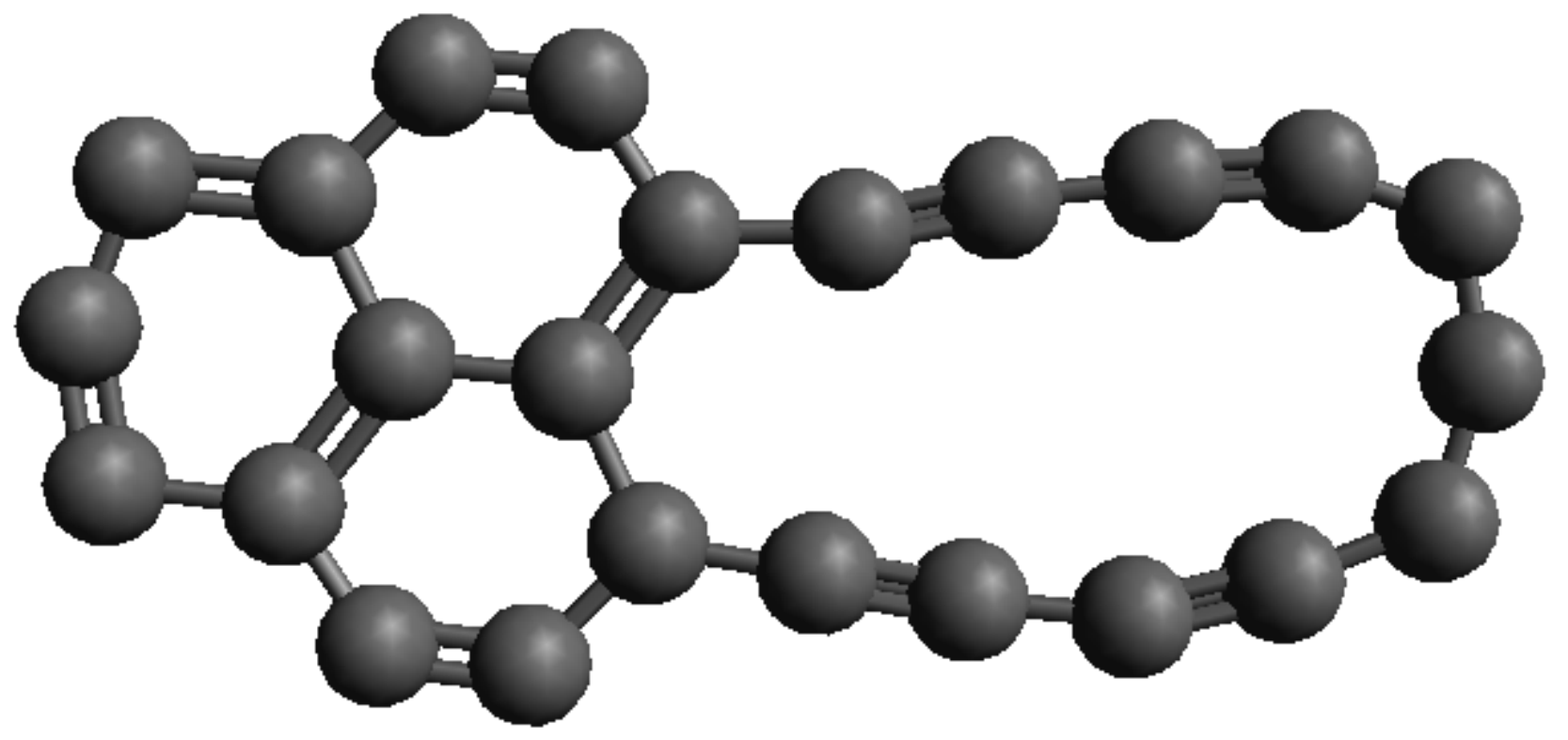}}
\caption{\label{fig:explode}An example of the ``exploding'' dPAH
 structures that result from optimizing the geometry of species with
 three (or more) linearly arranged benzenoids. }
\end{figure}

Stability issues also lowered the number of corresponding dPAHs. We
found that a PAH containing a linear chain of three or more benzenoids
generally caused the structure of the corresponding dPAH to
``explode'' (see e.g. Fig.~\ref{fig:explode} for an example). Again,
manual intervention was needed to detect and remove these species
since Gaussian considered these species to be converged
properly. These ``exploded dPAHs'' were not included in the
database. The final database produced for this paper thus contains
1550 PAHs and 805 dPAHs. The largest of which contains 34 carbon atoms.

\section{Content of the database: molecular structures}
\label{Sect:Structures}

\begin{table}[t]
\centering
\caption{Fractions of PAHs and dPAHs in the database with a given feature.}
\label{table:structtable}
\begin{tabular}{ccc}
\hline
Property & \% of total PAHs & \% of total dPAHs \\
\hline
Planar & 36 & 52 \\
Non-planar & 64 & 48 \\
Radical & 43 & 0 \\
Non-radical & 57 & 100 \\
Odd \#C & 43 & 46 \\
Even \#C & 57 & 54 \\
Pericondensed & 71 & 99 \\ 
Catacondensed & 29 & 1 \\
Hexagons only & 100 & 23 \\
Pentagon containing & 0 & 71 \\
Quadrilateral containing & 0 & 15 \\
\hline
\end{tabular}
\end{table}

\begin{figure*}
\resizebox{\hsize}{!}{%
  \includegraphics{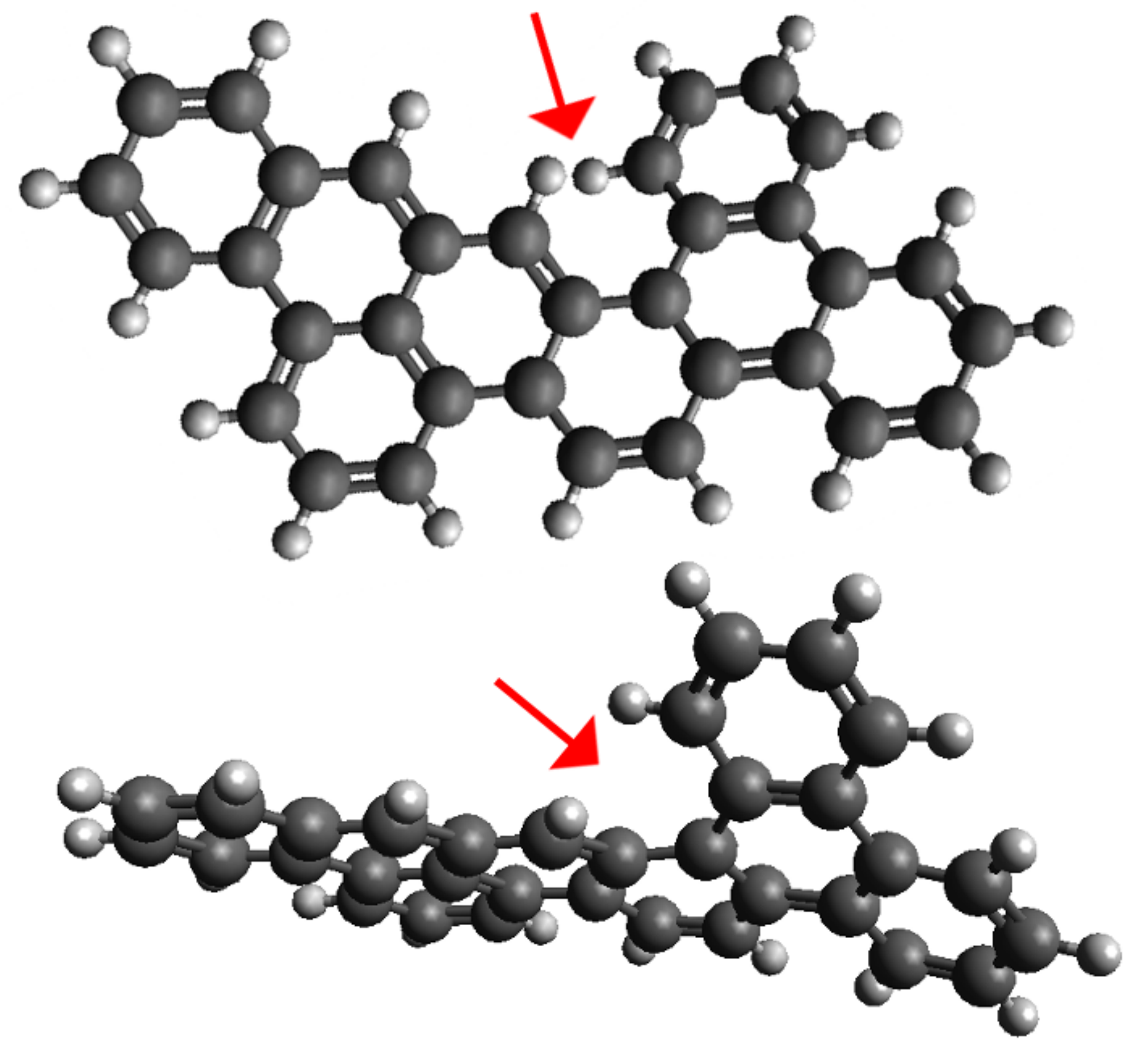}
  \includegraphics{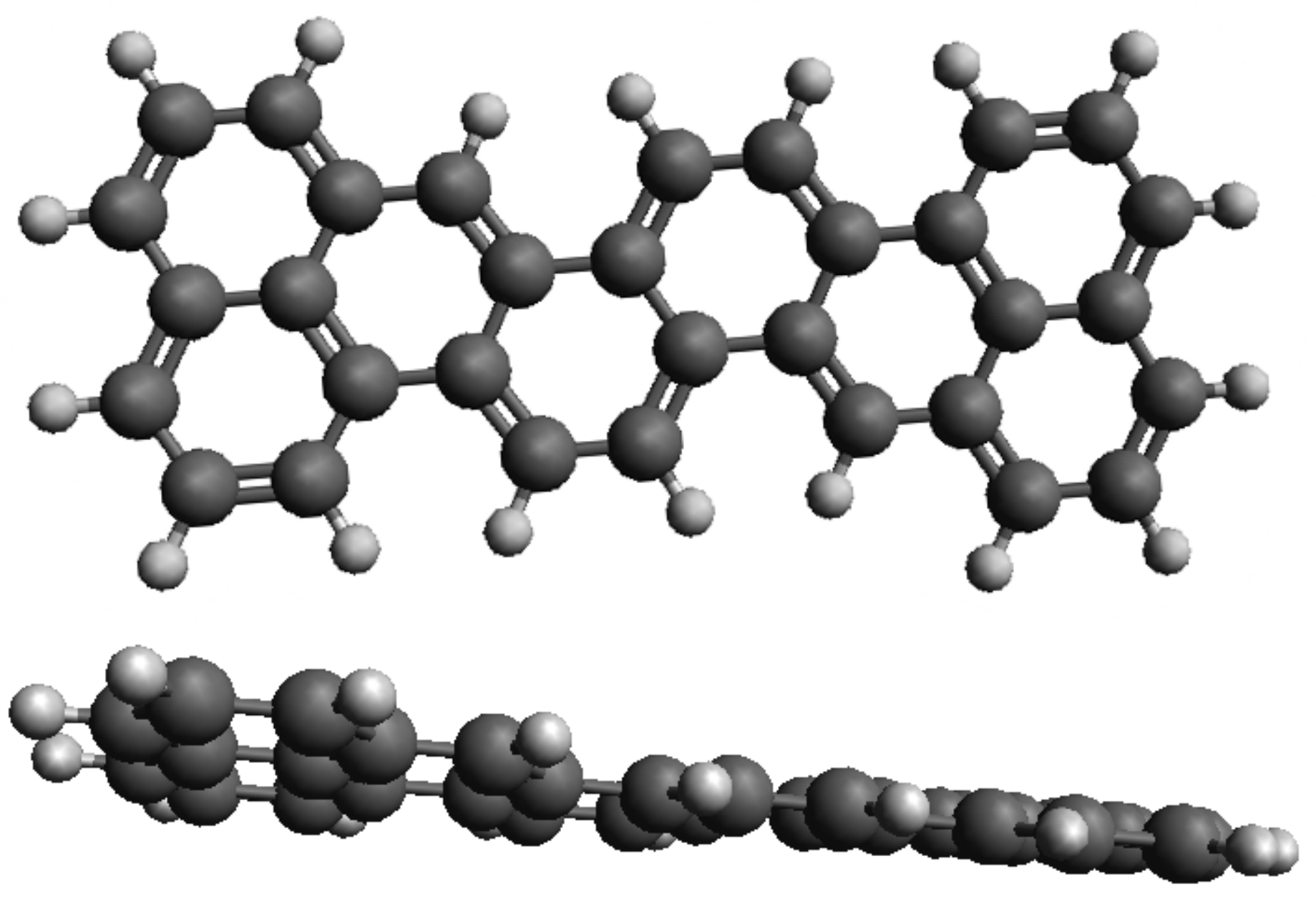}}
\caption{\label{fig:fjord}\label{fig:bridge}Two examples of
 deformations in PAH structures that lead to non-planarity. {\em
  (Left)} Top and side views of a ``fjord'' region (indicated by the red arrows) that results in a
 very strong deformation. {\em (right)} Top and side views of
 catacondensed bridges between pericondensed regions result in a
 slight twisting deformation.}
\end{figure*}

\begin{figure}
\resizebox{\hsize}{!}{\includegraphics{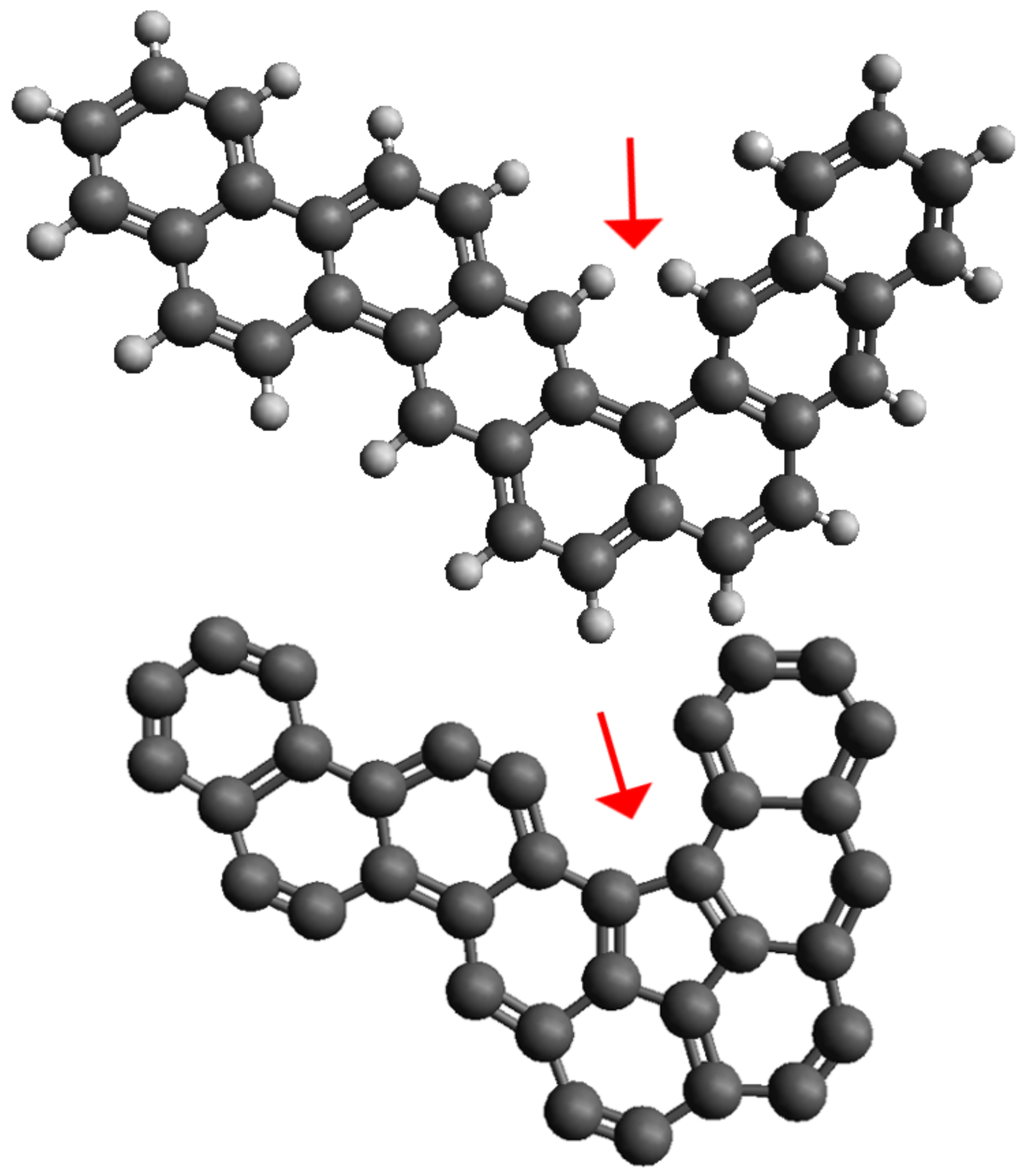}}
\caption{\label{fig:pent}An example of the formation of pentagonal
 rings upon dehydrogenation of PAHs with fjord regions. }
\end{figure}

Our complete and unbiased database of all possible small PAHs and
dPAHs (up to eight benzenoids) allows for a survey of the general
geometries and structures of the family of PAHs. We studied various
notable or interesting properties; a summary table of these results is
given in Table~\ref{table:structtable}.

In the process of optimizing the molecular geometries, we noticed that
deviations from a purely planar structure are very common for PAHs. In
fact, the majority of the PAHs and about half of the dPAHs were found
to be non-planar. This is somewhat contrary to the typical
representation in the literature of a PAH being represented as a
planar species. We recognized two different geometric configurations
that result in departures from planarity. First, we found that deep
bay regions or ``fjords'' (as shown in Fig.~\ref{fig:fjord}) are very
common; hydrogen-hydrogen repulsion in these fjords causes sometimes
strong deformations. A second configuration involves pericondensed
regions connected by catacondensed bridges, (see
Fig.~\ref{fig:bridge}) which causes a very slight twisting of the
molecule along the bridges. This slight twisting can be thought of as
the two pericondensed regions repulsing each other. It should also be
noted that some larger PAHs have multiple points of H-repulsion or
bridge twisting; this then introduces stereochemistry in larger PAHs.
For the work presented in this paper, it was assumed that the
trans-twisting was lower in energy than the cis-bending of the
geometries and therefore more important to the overall spectrum of a
species.

About half of the dPAHs also turn out to be non-planar. For the dPAHs
though, the departure from planarity is due to a different reason
since no hydrogen atoms are available to repulse anymore. For dPAHs
instead, it is the formation of pentagonal rings that causes the
non-planarity. A typical example is shown in Fig.~\ref{fig:pent}. When
removing the hydrogen atoms of a regular PAH with a fjord region, the
fjord region closes up into a pentagonal ring. The more these
pentagons were central to the molecular structure, the more bowl-like
the dPAHs become. Note though that the formation of a pentagon does
not always result in a non-planar species: as can be seen from
Table~\ref{table:structtable}, only 48\% of the dPAHs are non-planar,
whereas 71\% contain at least one pentagon. Quadrilateral shapes
(i.e. four-membered rings) also formed in some species, but always in
conjunction with pentagon formation.

It is also interesting to note from Table~\ref{table:structtable} that
there are very few ($\sim$1\%) catacondensed dPAH structures. This can
be explained by two reasons. First, as stated earlier, dPAHs with
linear structures are much less stable than clustered structures (see
also the example of exploding PAHs in Fig.~\ref{fig:explode}). The
second reason is more subtle. When removing the hydrogen atoms from a
catacondensed PAH species, often a pentagon forms which can turn the
structure into a pericondensed species.


\section{IR spectral properties of dPAHs}
\label{Sect:spectral_analysis}

\subsection{Calculating IR PAH and dPAH spectra}

Our DFT calculations result in the frequencies and intensities of the
normal modes of the considered species. To bring these frequencies in
better agreement with experimental values, we applied a scale factor
of 0.951, similar to the value used for many species in the NASA Ames
PAH database \citep[see e.g.]{2010ApJS..189..341B,PAHdbv2}.
Generally, these cannot be used to directly compare to astronomical
spectra. As mentioned before, the excitation of PAHs in an
astronomical context is through single-photon excitation followed by
fluorescent emission of IR photons that cools the molecule \citep[see
 e.g.][]{ATB89, Bakes_and_Tielens:1994, 2001ApJ...556..501B,
 2001ApJ...560..261B}. A realistic model spectrum for an astronomical
PAH would thus require detailed knowledge of the frequencies and
intensities of all transitions arising from vibrationally excited
states, in addition to the anharmonic couplings between the modes. In
practice, this information is not readily available for each
individual PAH species, and thus several approximations need to be
made. One such approximation is to use the {\em thermal approximation}
\citep[see e.g.][]{dHendecourt:thermalapprox} in which the state of
the molecule is described by its internal ``temperature'' -- the
average energy per vibrational mode. At any given temperature, the IR
emission scales with a Planck function at that temperature; in
essence, a full fluorescence calculation then follows the evolution of
the internal temperature and the corresponding emission over the
different IR active modes. Such calculations thus provide the scale
factors that have to be applied to the different band intensities to
account for how much time is spent at what temperature. One then
furthermore has to assume a line profile for each of the transitions
to obtain a full IR spectrum \citep[for a full description of such
 models, see the Appendix in ][]{PAHdbv2}.

A full cascade IR model calculation of the spectra of PAHs and dPAHs
would certainly also be the best for our purposes here. However, such
calculations are computationally expensive given the large number of
species in our database. Moreover, since our main purpose is to find
characteristic spectral feature for dPAHs, we are less concerned with
relative intensities than with peak positions. Thus, we followed
instead a much simpler approach and scaled the intensities by
multiplying them with a blackbody of a fixed temperature of
500~K. This corresponds to the assumption that the full fluorescence
cascade can be reproduced by a single ``characteristic'' temperature,
and that that temperature is the same for all species. The latter
assumption most certainly does not hold: indeed, smaller PAHs have
less vibrational modes and thus a higher temperature (energy per mode)
for a given UV absorption than larger PAHs. Thus, our approach will
somewhat underestimate the emission of smaller PAHs at short
wavelengths. For our purposes here, we are mainly interested in the
frequencies of the features rather than their intensities; therefore,
this is an acceptable default. A gaussian profile with a linewidth of
$5cm^{-1}$ was chosen.

\subsection{Spectral characteristics of dPAHs}

\begin{figure}[t]
\resizebox{\hsize}{!}{\includegraphics{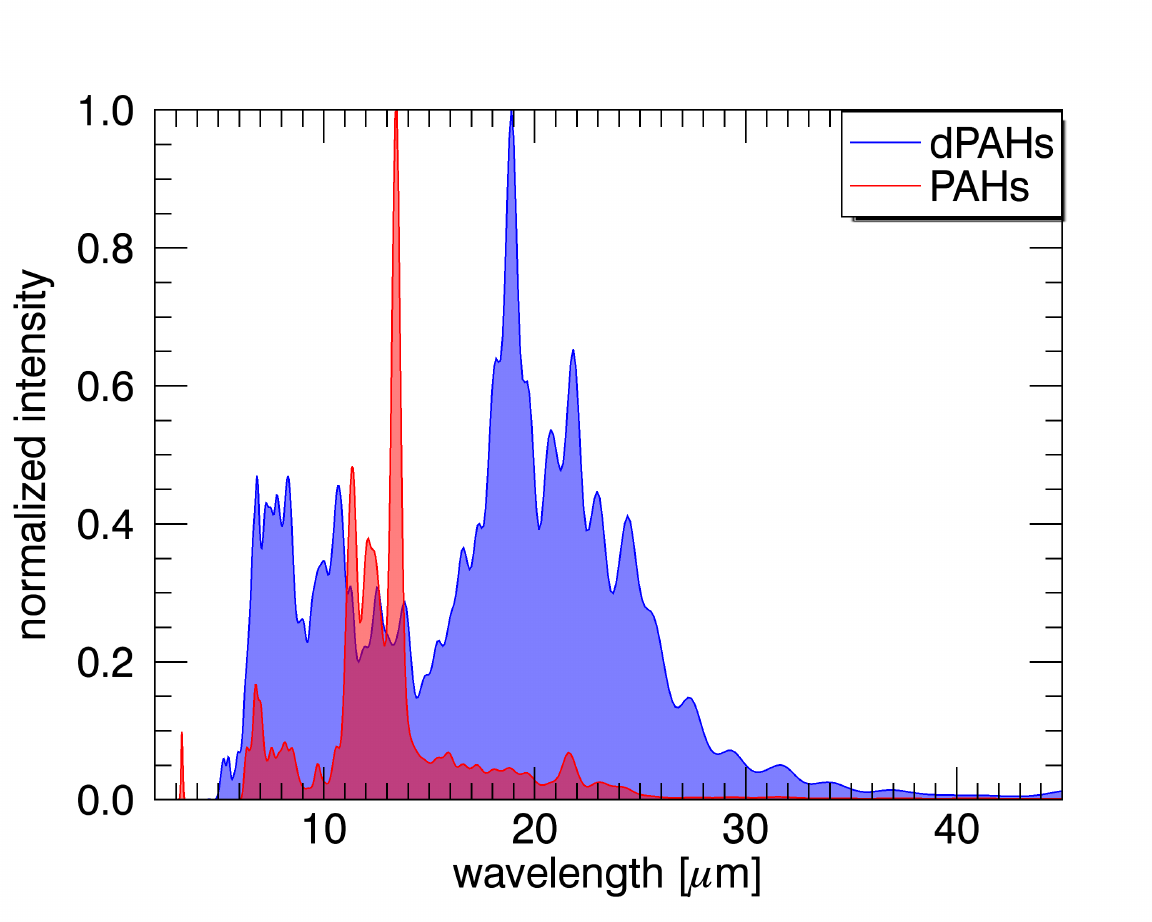}}
\caption{\label{fig:averagespec}The average spectrum of all PAHs in our
 database (red) compared to the average spectrum of all dPAHs
 (blue).}
\end{figure}

\begin{figure}[t]
\resizebox{\hsize}{!}{\includegraphics{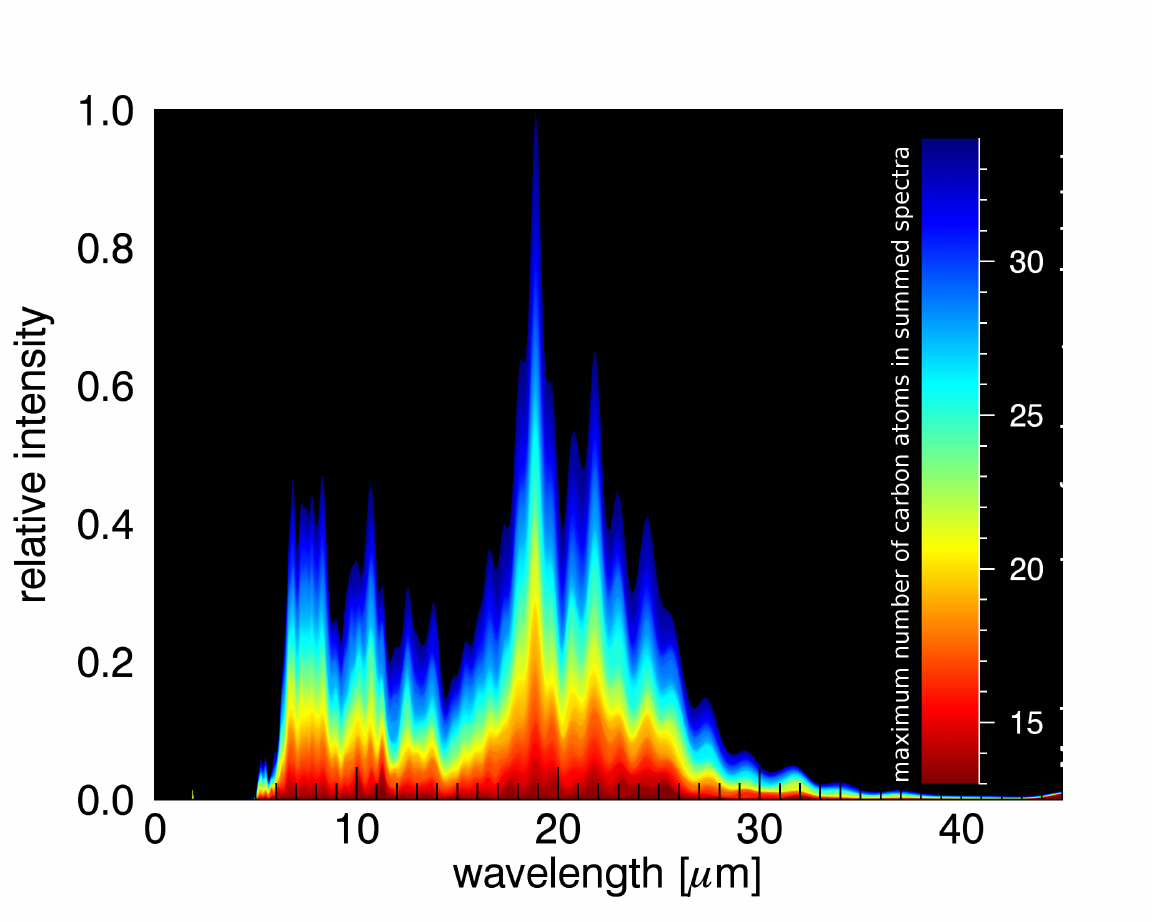}}
\caption{\label{fig:fireplot}Cumulative spectra of dPAHs as a function
 of size. At the bottom are the spectra of the smallest species
 (color coded red); we cumulatively added the spectra of ever larger
 species and changed the color to blue. Careful analysis of the
 features in this plot can discriminate between features that are
 common to all dPAHs and features that originate mainly from smaller
 or larger species.}
\end{figure}

We now turn our attention to investigating the typical spectral
features of dPAHs. To characterize the features that distinguish a
typical dPAH spectrum from a PAH spectrum, we first produced the
average spectrum (arithmetic mean) of all PAHs and that of all dPAHs,
and visually compared them for any features of interest (see
Fig.~\ref{fig:averagespec}). To further distinguish how size might
influence spectral features, we also produced color-coded cumulative
spectra in Fig.~\ref{fig:fireplot}.

From a close comparison of the average PAH and dPAH spectra, we find
notable dPAH features at 5.2$\mu$m, 5.5$\mu$m, and 10.6$\mu$m, as well
as a broad, structured emission forest from $\sim$16--30$\mu$m with its
strongest feature at 19$\mu$m. These features are absent
in the average PAH spectrum, and are thus a first indication that they
may represent diagnostic features to recognize a population of dPAHs
in astronomical spectra. Our findings compare well to the results by
\citet{charlie:dehydrogenated} who also found a 5.25$\mu$m feature and
emission in the 15-25$\mu$m region they attribute to neutral (large) dPAHs.

It is interesting to note that features of the forest (i.e. emission between 15-30$\mu$m) appear to follow
a broad, somewhat gaussian-like distribution centered around
20$\mu$m. Note however that individual species do not show this
distribution but have a small number of discrete bands. From
Fig.~\ref{fig:fireplot} it can be seen that as larger species are
added to the average spectrum, some of the features in this forest
grow in prominence and are not washed out. This suggests that even
some of the features in this forest may have diagnostic power. Another
interesting point is that the strongest feature of the forest is at
19$\mu$m; from Fig.~\ref{fig:fireplot}, it furthermore appears that
this feature becomes more prominent than the other forest features 
for the larger dPAHs. In recent
years, emission at this wavelength in astronomical observations has
been attributed to fullerenes, but our work shows that a feature at
that wavelength may also be due to a family of dPAH-like
species. 

\begin{figure}[t]
\resizebox{\hsize}{!}{\includegraphics{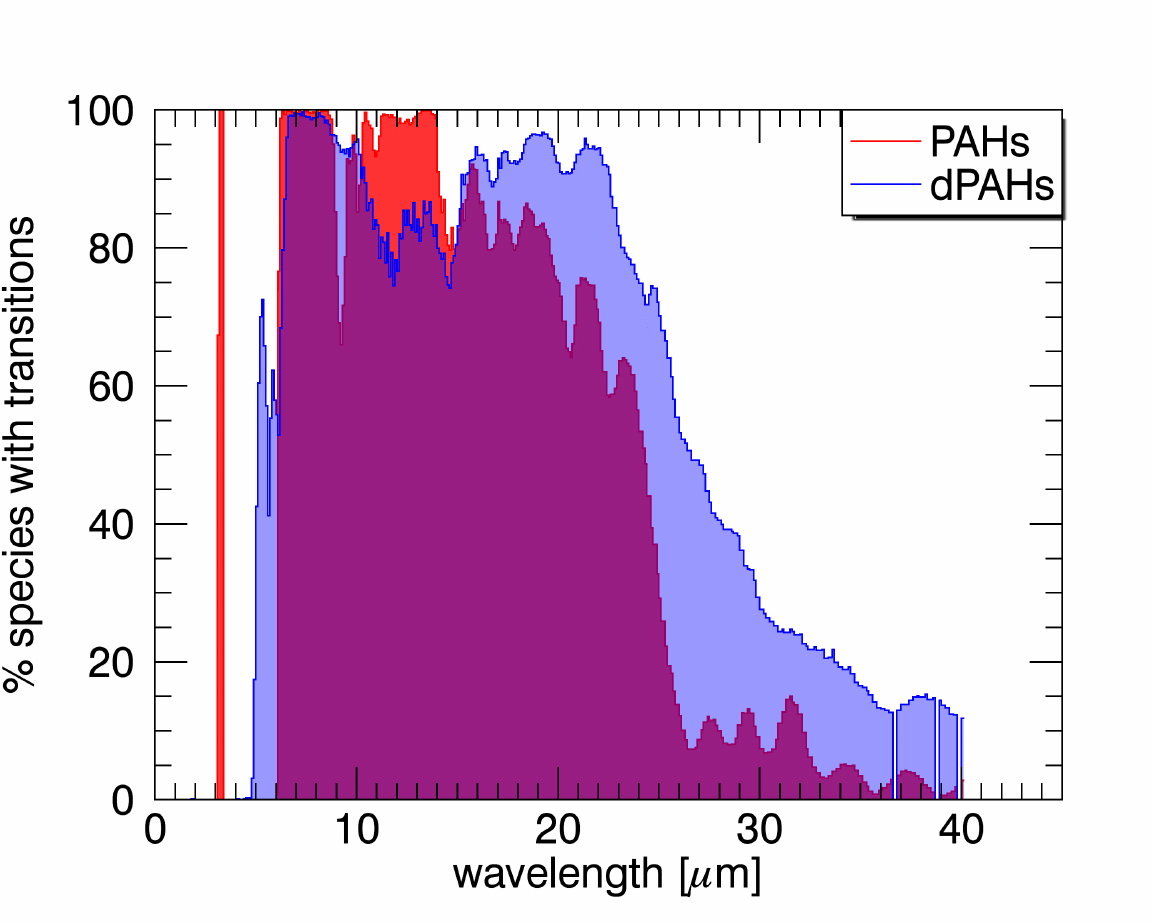}}
\resizebox{\hsize}{!}{\includegraphics{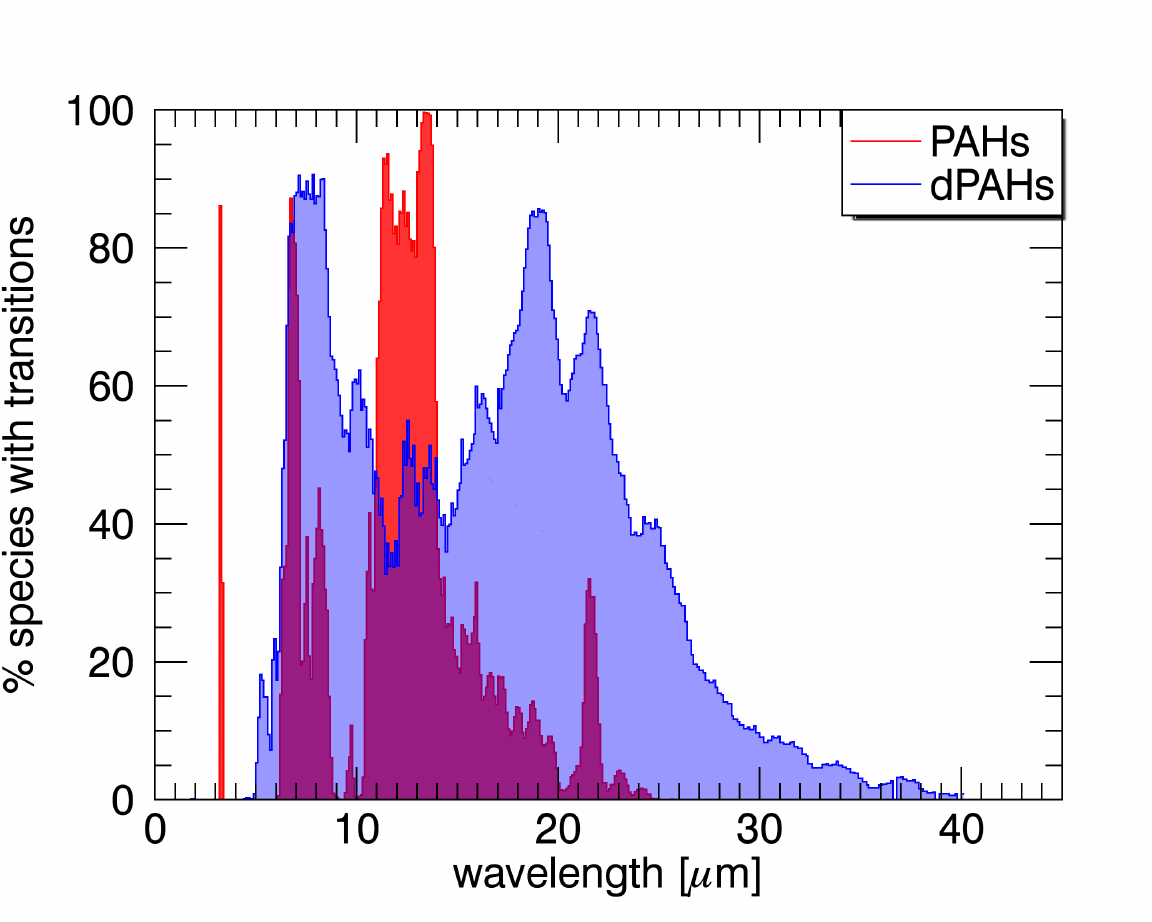}}
\resizebox{\hsize}{!}{\includegraphics{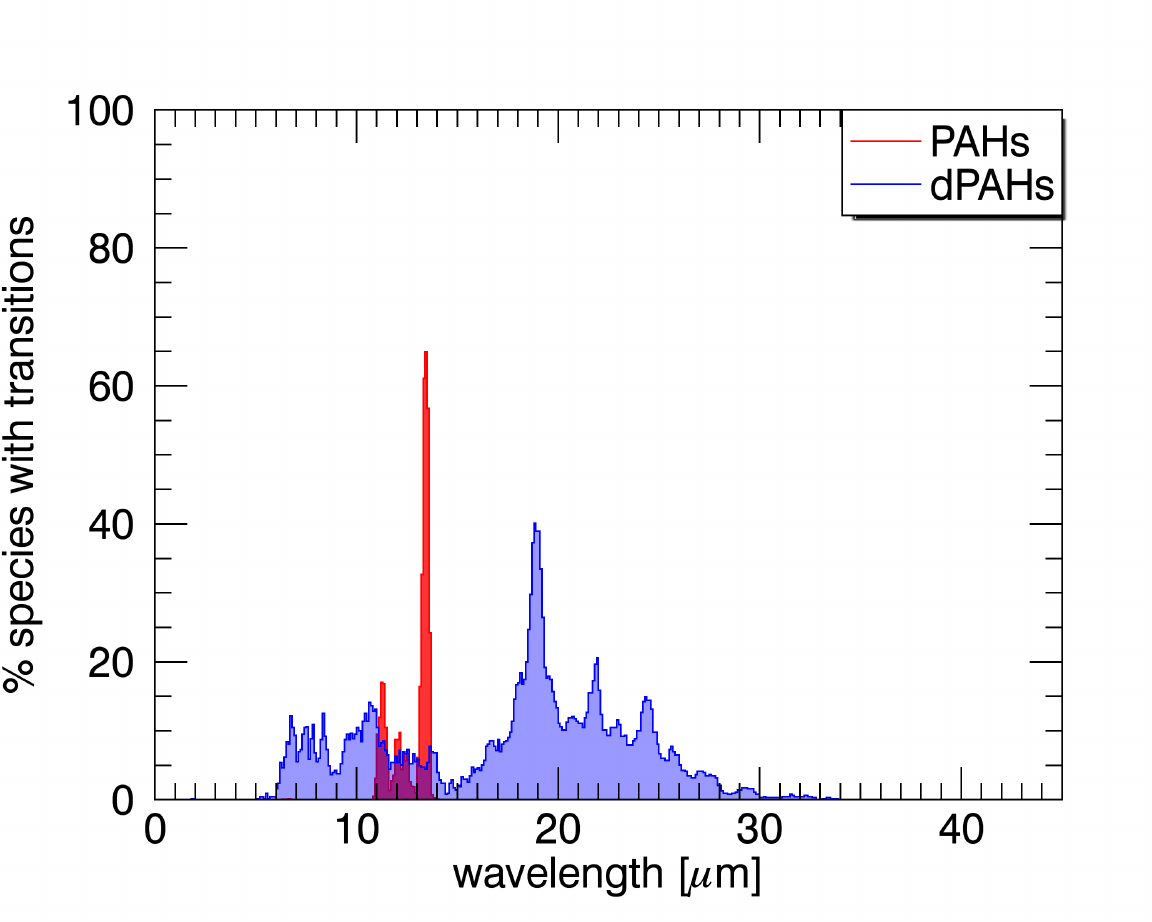}}
\caption{\label{fig:histo}Histograms representing the occurrence
 frequency of transition for PAHs and dPAHs in our database. {\em
  (top)} All transitions -- essentially no intensity cut-off is
 applied. {\em (middle)} Medium and strong transitions only. {\em
  (bottom)} Strong transitions only.}
  \end{figure}

One could argue that the average dPAH spectrum from our database may
not be a good representation of what could be expected in
space. Indeed, physical conditions may favour a subclass of species
with spectral properties that are somewhat different from the
average. Thus, it may be more appropriate to consider as a diagnostic
the transitions that are {\em most common} among species, i.e. that
occur most frequently in our database. We thus studied the transition
recurrence by creating histograms of the fraction of species with a
transition at a given wavelength. We did this for both PAH and dPAH
species, and used a bin size of 0.1$\mu$m. To distinguish between weak
and strong transitions, we further applied three different intensity
cut-offs to our list of transitions before creating the histogram. The
three histograms we thus obtained for PAHs and dPAHs are shown in
Fig.~\ref{fig:histo}.

The top histogram in Fig.~\ref{fig:histo} represents essentially no
cut off and shows that both PAHs and dPAHs have frequent transitions
over the entire wavelength range. There are three clear differences
overall between PAHs and dPAHs: {\em (i)} all PAHs have a transition
at 3.3$\mu$m corresponding to the C-H stretching mode; understandably,
none of the dPAHs shows this transition. {\em (ii)} dPAHs show
transitions around 5.5$\mu$m where none of the PAHs have
transitions. In fact, it turns out that {\em all} of the dPAH species
have at least one transition near 5.5$\mu$m. {\em (iii)} dPAHs have
many more transitions at the longer wavelengths; this is notable from
about 16$\mu$m onward. 

When removing the weakest transitions from the database by increasing
the intensity cut-off (middle histogram), these differences between
PAHs and dPAHs remain and the long-wavelength dPAH transitions are
even more pronounced. Additionally, two clear regions become apparent
where many dPAHs have medium or strong transitions, but the regular
PAHs don't: an emission feature at 10.6$\mu$m and a second one at
19$\mu$m. Keeping only the strongest transitions (bottom histogram) we
find a histogram that is very similar to the average dPAH spectrum in
Fig.~\ref{fig:averagespec}, but not so for the PAHs. The regions of the typical dPAH
spectra that were shown earlier to stand out from a typical PAH spectra also
remain present.
\\

Thus, both from a comparison of the average PAH and dPAH spectra, and
from the histogram analysis we reach the same conclusion: we would
expect that a population of (small) dPAHs would be spectroscopically
detectable by their features at 5.5$\mu m$, 10.6$\mu m$, 19$\mu m$ and
a continuum-like 'forest' of individual features roughly between
16-30$\mu m$.

\subsection{Correlation analysis}

\begin{figure}[t]
\resizebox{\hsize}{!}{\includegraphics{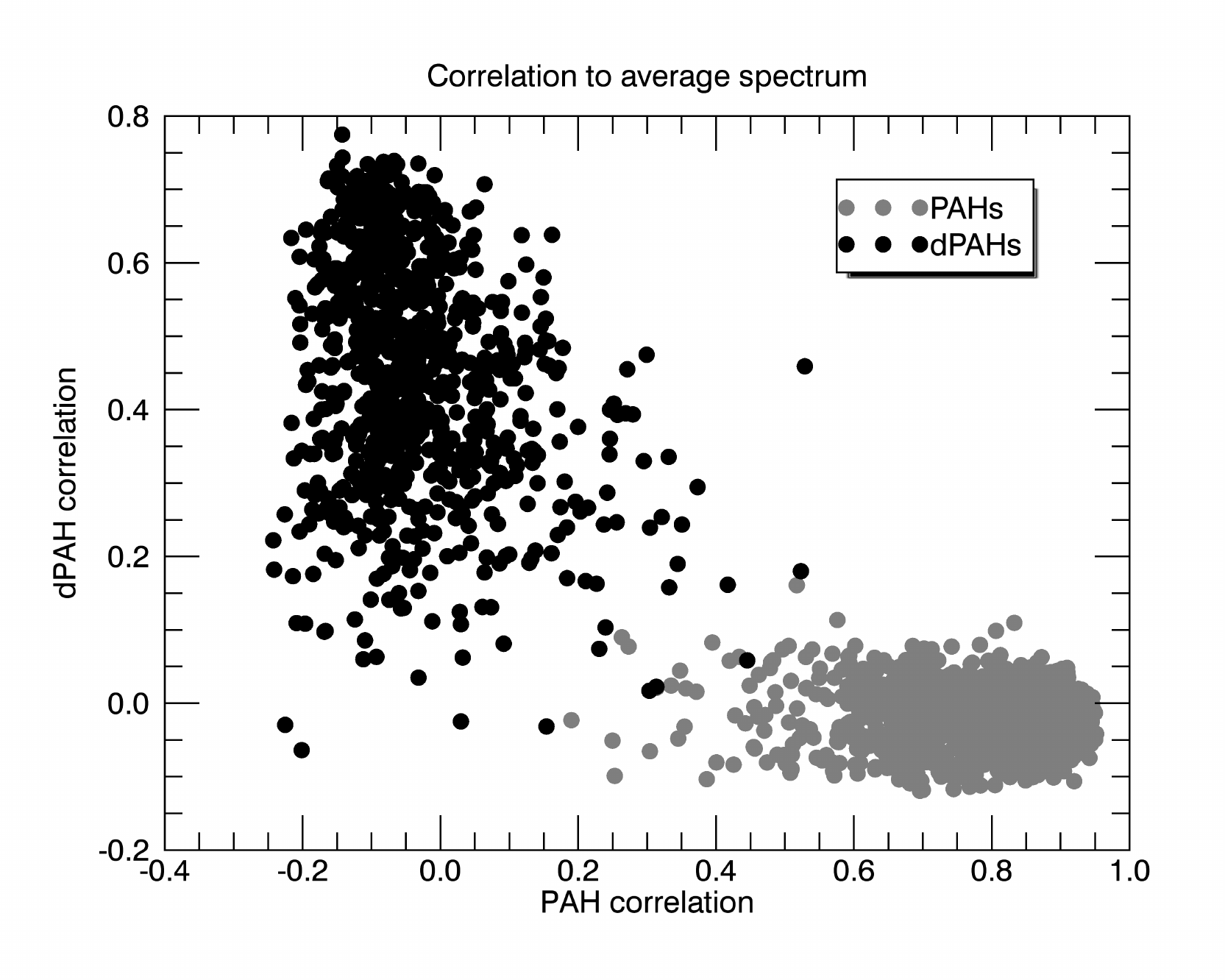}}
\caption{\label{fig:correlation}The linear Pearson correlation
 coefficients of individual PAH (gray) and dPAH (black) spectra with
 respectively the average PAH spectrum (horizontal axis) and average
 dPAH spectrum (vertical axis). }
\end{figure}

We also addressed the question how ``characteristic'' the average dPAH
spectrum is. To this end, we calculated the linear Pearson correlation
coefficient between each of the individual PAH spectra and the
average PAH and dPAH spectra, and a similar exercise was carried out for the
individual dPAH spectra. We calculated the correlation coefficient
over the range 5--20$\mu$m in order to exclude some of the most
obvious differences (such as the 3.3$\mu$m PAH feature and much of the
16-30$\mu$m dPAH features). The resulting correlation coefficients are
shown in Fig.~\ref{fig:correlation}.

We found that generally, individual PAH spectra correlate rather well
with the average PAH spectrum with an average correlation coefficient
of 0.74, with only a fairly modest spread (the majority lies between
0.6 and 0.95). Thus, most individual PAH spectra look fairly
similar. This is not so for the dPAH spectra with an average
correlation coefficient of only 0.37 between the individual dPAH
spectra and the average dPAH spectrum. Moreover, there is a much
larger spread (with most dPAHs having a correlation coefficient
roughly between 0.2 and 0.75). Thus, dPAH spectra tend to be much more
different from one species to the next. Consequently, it is much
harder to ``characterize'' a typical dPAH spectrum. Referring back to
the molecular structures of PAHs and dPAHs this difference can be
attributed to the fact that while PAHs tend to be constructed of
hexagons that deviate very little from each other, dPAHs are
constructed of deformed hexagons that differ greatly from one species
to the next, as well as internally to a given species. This effect is
most prominent in the lower energy transitions, and is likely the
cause of the 16-30$\mu m$ continuum. It was found that each species has an almost
unique signature in this region. But clearly, these deformations leave
their mark in the C-C stretch region as well. These features could provide clues as to
which species are present in space, and possible lead to a detection of a single species.
The dPAH and PAH spectra created for this paper will soon be available in the 
NASA AMES PAH database for others to make use of. 

Finally, we considered the correlations between individual PAH spectra
and the average dPAH spectra, and vice versa. It is immediately clear
from Fig.~\ref{fig:correlation} that the two groups of spectra are
very different. Indeed, the correlation plot clearly shows two
clusters: one corresponds to the PAH spectra; the other to the dPAH
spectra. None of the PAH spectra correlates well with the average dPAH
spectrum, and similarly none of the dPAHs correlates well with the
average PAH spectrum. Thus, PAHs and dPAHs are spectroscopically very
distinct!

We carried out a similar analysis for the other subclasses in
table \ref{table:structtable}; however, we could not find any
significant spectroscopic diagnostic for any of those other
subclasses.


\section{A search for interstellar dPAHs}
\label{Sec:NGC7023}

\begin{figure}[t]
\resizebox{\hsize}{!}{\includegraphics{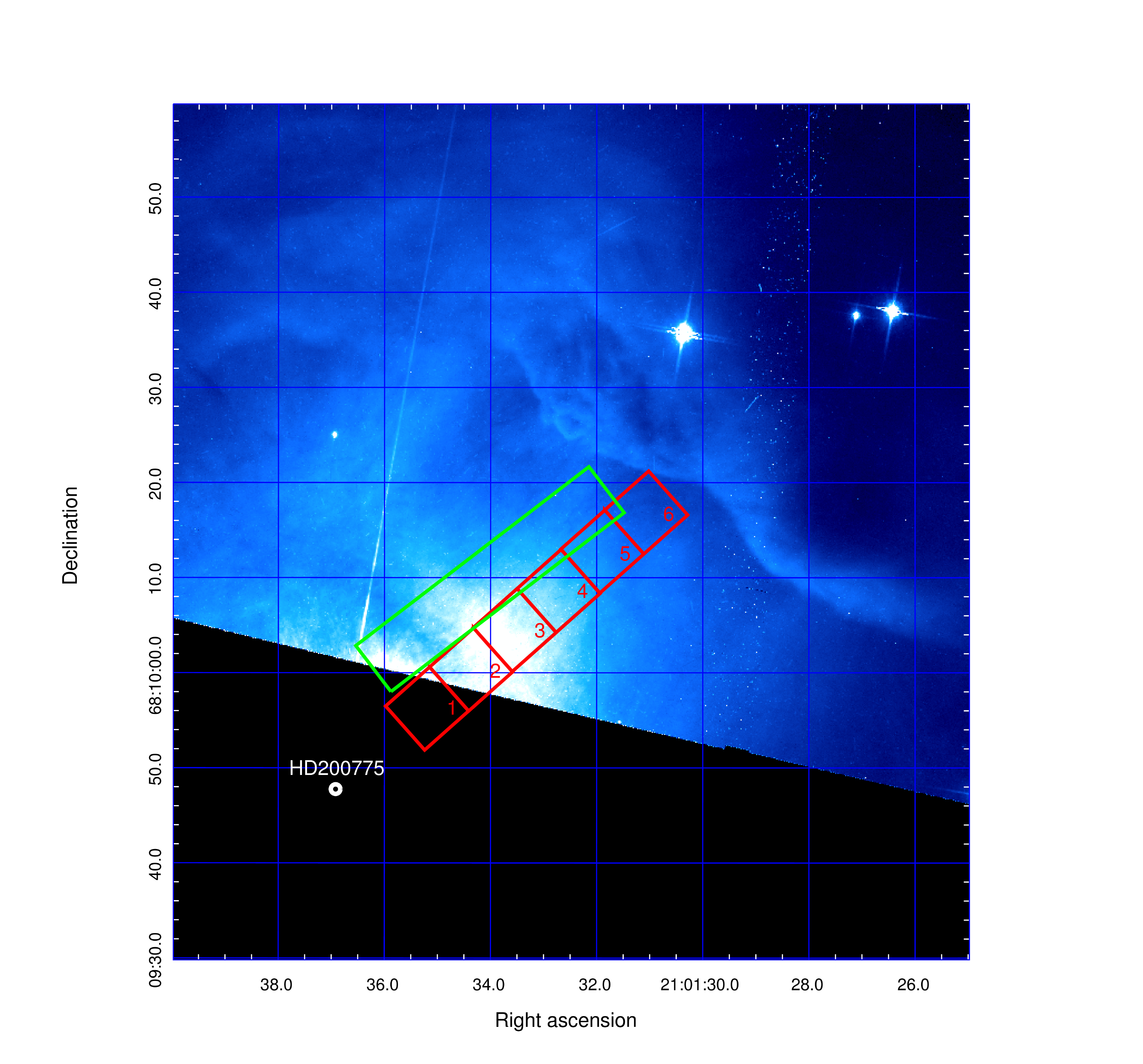}}
\caption{\label{fig:overobs} A HST image of NGC7023 with the apertures used in 
this paper (red boxes) and the extraction region used by 
\citet{2012PNAS..109..401B}. The illuminating star, HD200775, is indicated by a circle. }
\end{figure}

\begin{figure*}[t]
\resizebox{\hsize}{!}{\includegraphics{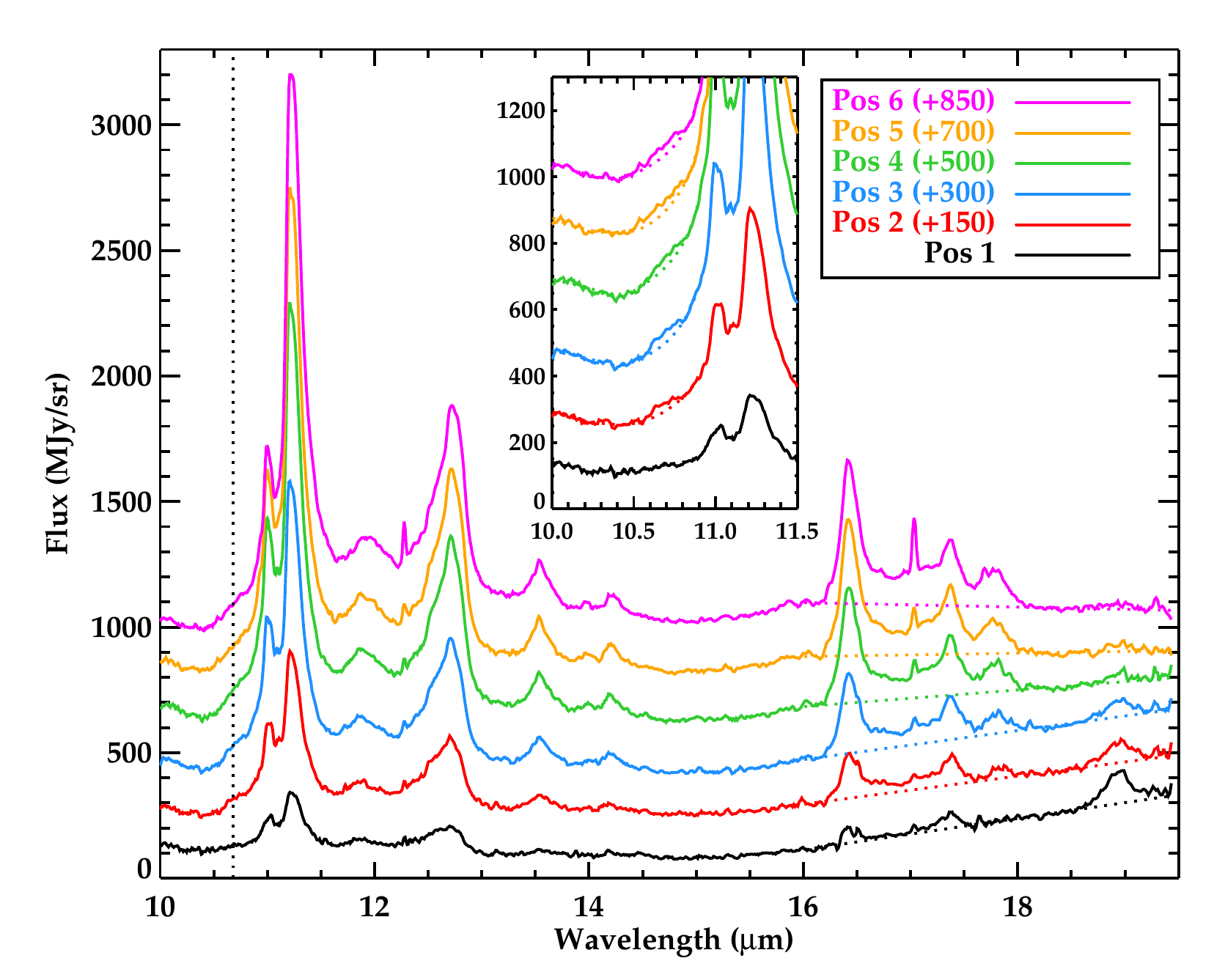}}
\caption{\label{fig:sixspec}The six Spitzer-IRS SH spectra for
 NGC~7023 as a function of increasing distance to the illuminating
 star, corresponding to the apertures shown in
 Fig.~\ref{fig:overobs}. We smoothed the spectra with a 3 pixel wide
 boxcar average and we have offset them for clarity using the offsets
 indicated in the legend. The dotted lines redward of 16$\mu$m are
 least absolute deviation fits to the local continuum for each
 spectrum. The dotted line at 10.68$\mu$m marks the center of a very
 weak emission feature (best seen in positions 2 and 3) in the wing
 of the 11.0 $\mu$m PAH band. The inset shows a close-up of this
 wavelength range, where the dotted lines now represent a spline fit
 through the wing of the 11.0$\mu$m PAH band, used to extract the
 much weaker 10.68$\mu$m feature (see text for details). }
\end{figure*}

Our analysis above suggests that if a population of dPAHs is present
in interstellar environments, we might be able to detect their
characteristic spectral features near 5.5$\mu$m, 10.6$\mu$m and
19$\mu$m, and as structured continuum emission at longer
wavelengths. It is very difficult in the best case to establish the
latter type of emission. At the same time, a feature at 19$\mu$m could
also be due to fullerenes. At the shorter wavelengths, PAH features
are known at 5.25 and 5.7 $\mu$m and those have been attributed to
combination bands \citep{2009ApJ...690.1208B}. Thus, it seems that the
feature near 10.6$\mu$m might offer the best prospects to search for
dPAHs. Note however that also here, there might be some confusion
since \citet{2013A&A...550L...4B} recently reported a feature at
10.53$\mu$m (with a FWHM of 0.11$\mu$m) in a spectrum of NGC~7023 that
they attribute to emission of C$_{60}^+$.

NGC~7023 is in fact an ideal object to search for dPAH features for a
variety of reasons. A large amount of high-quality IR observations is
available for this object (see below), and these data have already
been studied in detail in terms of PAHs and fullerenes \citep[see
 e.g][]{2007ApJ...659.1338S, 2010ApJ...722L..54S,
 2012PNAS..109..401B, 2013A&A...550L...4B, Boersma:NGC7023}. Of
particular interest here is the work by \citet{2012PNAS..109..401B}
who studied variations in the IR spectra as one moves from the PDR to
regions much closer to the illuminating star \object{HD 200775}. In
the PDR, far away from the star, the spectrum shows very strong
emission from PAHs and no evidence for fullerenes. Approaching the
central star though, \citet{2012PNAS..109..401B} find that the
abundance of PAHs decreases, while that of C$_{60}$ increases. They
argue that fullerenes form from large PAHs ($\sim$70 C atoms) that
become dehydrogenated by UV radiation, subsequently close up and
shrink down to become C$_{60}$. Thus, large dPAHs are explicitly
suggested as an intermediate step in the formation of
fullerenes. However, smaller PAHs could also become fully
dehydrogenated at even lower doses of UV radiation. Thus, dPAHs may be
formed over a much larger area than only between the PAHs and the
fullerenes. If enough of these dPAHs survive for long enough, they
might be detectable.

A large amount of IR spectroscopic observations of NGC~7023 were
carried out by the IRS spectrograph \citep{Houck:IRS} on board the
Spitzer Space Telescope \citep{Werner:spitzer}. Here, we use primarily
the high-resolution spectra taken with the Short-High (SH) module
covering the wavelength range 10--19$\mu$m at a resolving power of
$R\sim600$ (AOR 3871232). We will also include some discussion about the
low-resolution spectra taken with the Short-Low (SL) module covering
the wavelength range 5.2--14$\mu$m(AORs 3871488 and 3871744). We did 
not apply background subtraction. We cleaned the data using {\it cubism}'s 
automatic bad pixel generation with $\sigma_{TRIM} = 7$ and Minbad-fraction 
= 0.5 and 0.75 for the global bad pixels and record bad pixels. We have 
extracted spectra corresponding to 6 different
positions throughout the nebula at increasing distances from the
illuminating star (see Fig.~\ref{fig:overobs}). The chosen positions
are comparable to those by \citet[][their extraction regions are also shown in
 Fig.~\ref{fig:overobs}]{2012PNAS..109..401B}.

\begin{figure}[t]
\resizebox{\hsize}{!}{\includegraphics{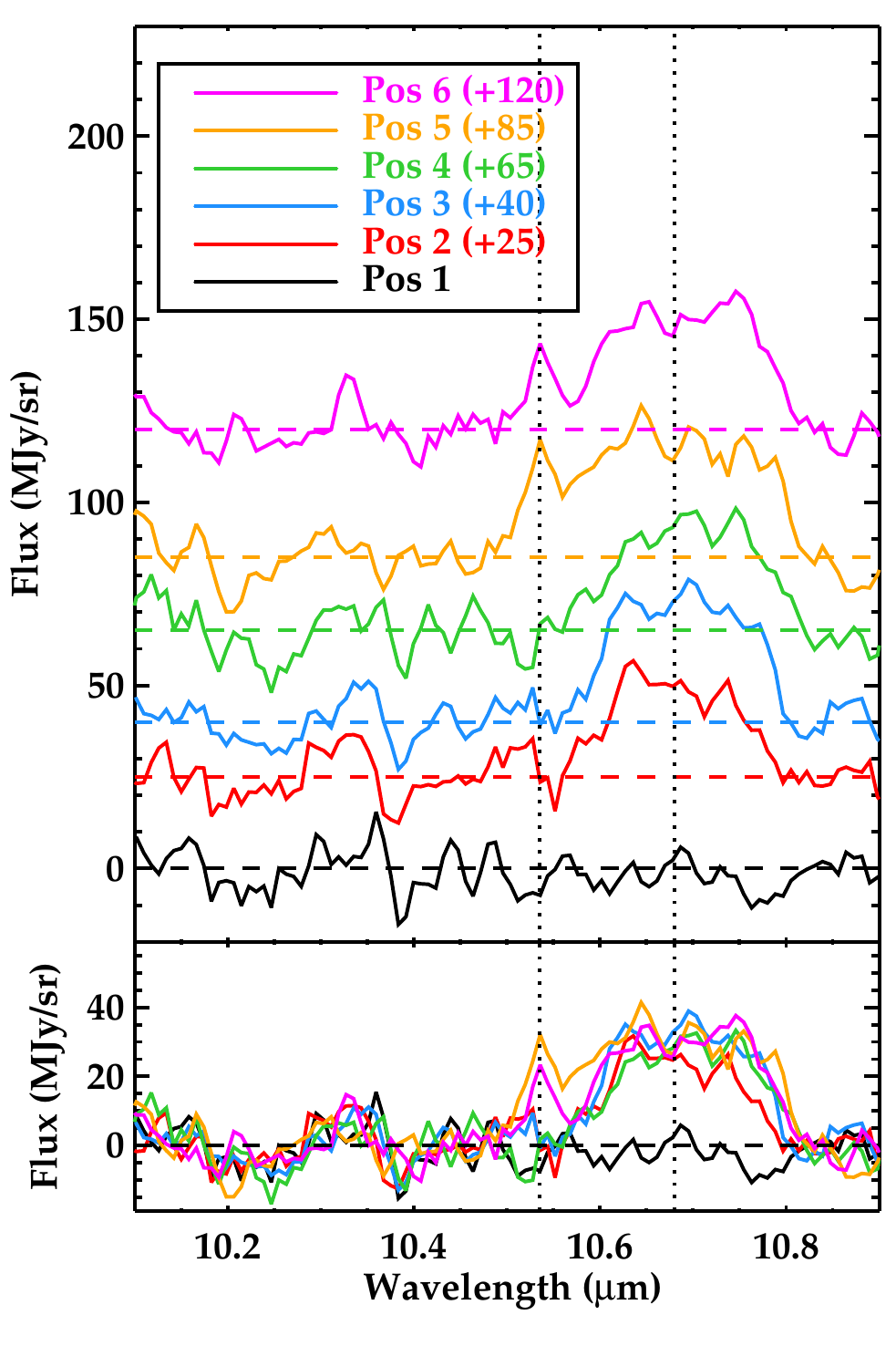}}
\caption{\label{fig:10.68um}The spectra of NGC~7023 (smoothed as in
 Fig.~\ref{fig:sixspec}, and offset) after subtracting the cubic
 spline fit to represent the wing of the 11.0$\mu$m feature (dotted
 line in the inset of Fig.~\ref{fig:sixspec}). The feature centered
 at 10.68$\mu$m (indicated by the right dotted line) is clearly present in
 positions 2--6, but absent in position 1. An additional, much
 narrower emission feature centered at 10.53$\mu$m (indicated by the left
 dotted line) appears in positions 5 and 6. The dashed lines indicate the zero-point level for each spectrum.}
\end{figure}

\begin{figure}[t]
\resizebox{\hsize}{!}{\includegraphics{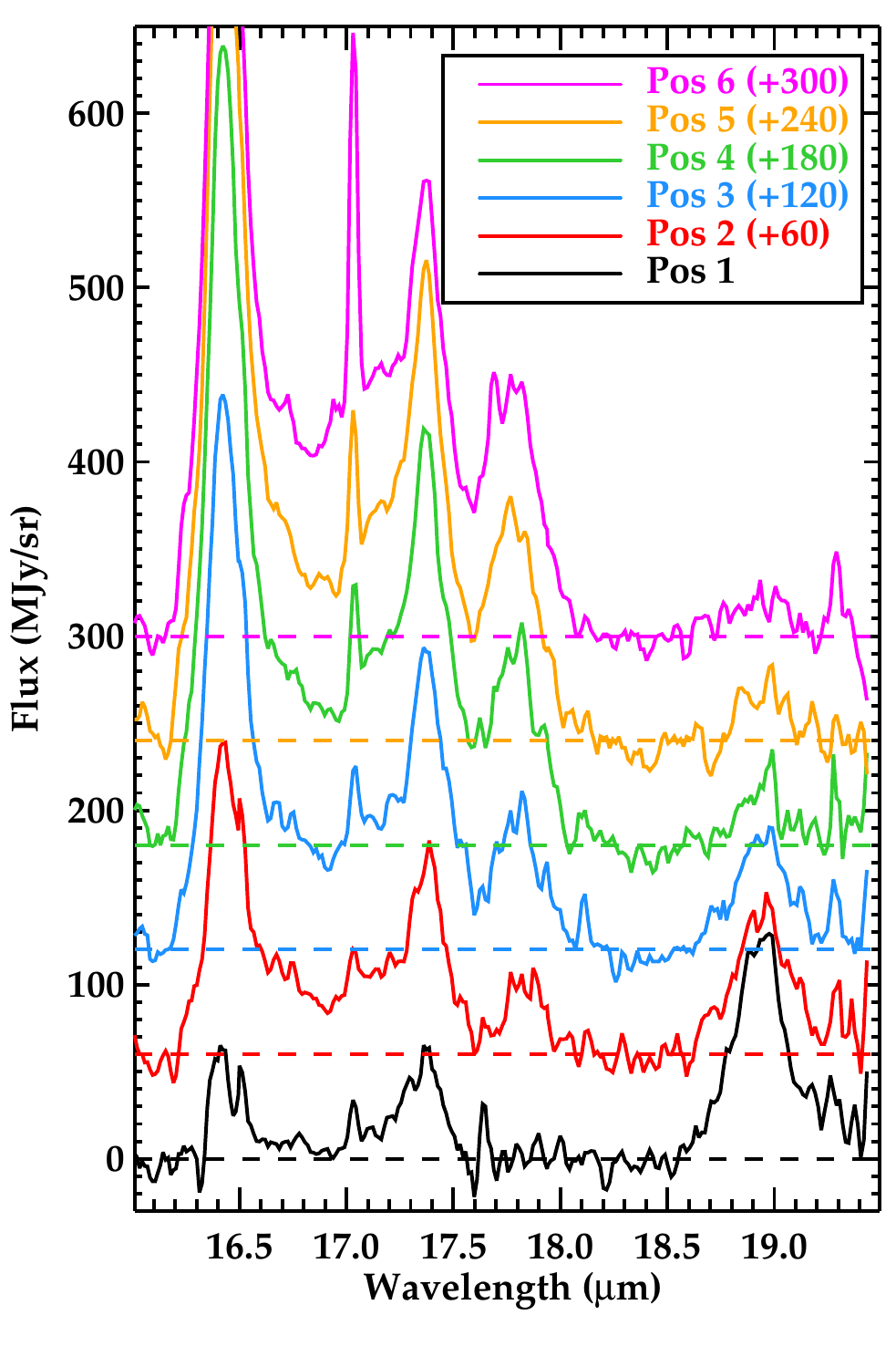}}
\caption{\label{fig:19um}The spectra of NGC~7023 (smoothed as in
 Fig.~\ref{fig:sixspec}, and offset) after subtracting the linear fit
 to the continuum (the dotted lines in Fig.~\ref{fig:sixspec}). The
 dashed lines indicate the zero-point level for each spectrum. }
\end{figure}

The full 10--19$\mu$m spectra (Fig.~\ref{fig:sixspec}) reveal the same
overall spectral variations that were described by
\citet{2012PNAS..109..401B}: strongest PAH emission in position 6
(furthest away from the illuminating star), and gradually less PAHs
and more fullerenes (17.4$\mu$m and 18.9$\mu$m bands) toward position
1. Close inspection of this figure reveals a very weak feature near
the predicted wavelength of 10.6$\mu$m (centered at 10.68$\mu$m,
indicated with the dotted line), most clearly seen in the spectra
corresponding to positions 2 and 3. To extract this feature, we need
to somehow estimate the ``local continuum'', i.e. the underlying
contribution of the wing of the 11.0$\mu$m feature. To do so, we first
defined a set of 7 anchor points in the spectrum of position 2 between
10.0 and 10.95$\mu$m, but excluding the range 10.5--10.8$\mu$m where
the feature is located. A cubic spline through these anchor points
then represents an estimate of the local continuum. We then fixed the
wavelengths of these anchor points but iteratively allowed their flux
values to vary to find the cubic spline that provides the best fit to
the spectrum in this range (but still excluding the 10.5--10.8$\mu$m
range). We then did the same for the remaining positions, always using
the same 7 fixed wavelengths as anchor points. The resulting curves are
shown as dotted lines in the inset of Fig.~\ref{fig:sixspec}. They are
smooth, contain no spectral structure, and seem to correspond well to
what might be the wing of the 11.0$\mu$m feature. We are thus
confident that this is a reasonable approximation to the spectrum
under the weak 10.68$\mu$m feature. However, we realize that there is
a fair amount of uncertainty in this determination.

Fig.~\ref{fig:10.68um} shows the resulting 10.1--10.9$\mu$m spectra
after subtraction of this local continuum. The feature at 10.68$\mu$m
is now very clearly visible in the spectra of positions 2--6, but
absent in position 1. It is also interesting to note that in positions
5 and 6, an additional, even weaker emission feature is superposed on
the spectrum, with a central wavelength of 10.53$\mu$m. This coincides
with the reported wavelength of a weak feature (with a peak height of
$\sim$6 MJy sr$^{-1}$) that was associated with C$_{60}^+$ by
\citet{2013A&A...550L...4B} in a spectrum of NGC~7023 very close to
the central star. We do not find evidence for such a feature in our
position 1 spectrum (where we would expect it) but surprisingly at the
positions furthest away from the star, and much stronger than the
feature reported by \citet{2013A&A...550L...4B}. It is also slightly
narrower than the reported C$_{60}^+$ feature. Where the 10.68$\mu$m
feature is present, it appears roughly the same in terms of width,
shape and overall (absolute) strength. In contrast, the intensity of
the PAH emission increases by about a factor of 10 between positions 1
and 6 (see Fig.~\ref{fig:sixspec}). The width of the feature (FWHM of
about 0.2$\mu$m) is comparable to e.g. the 11.2 or 16.4 $\mu$m PAH
features. It it thus possible that the 10.68$\mu$m is indeed the weak
emission of a population of dPAHs.

If dPAHs are responsible for the 10.68$\mu$m feature, they should also
emit at other wavelengths. We first considered the 19$\mu$m region. It
is clear from Fig.~\ref{fig:sixspec} that a well-defined emission band
is present in the spectra of positions 1--3; there is a hint though of
weaker emission in the other positions as well. To better evaluate
this possibility, we fitted a least-absolute-deviation straight line
to the continuum longward of 16$\mu$m, excluding the wavelength ranges
of PAH and C$_{60}$ features (see Fig.~\ref{fig:sixspec}). While this
may not the best possible representation of the local continuum, it
ensures that we do not introduce any artificial features in the
spectrum. Fig.~\ref{fig:19um} shows the resulting continuum-subtracted
spectra in this wavelength range. This figure now more convincingly
shows that weak excess emission at 19$\mu$m also appears at the other
positions. While a 19$\mu$m feature could be due to C$_{60}$, this is
not likely the case for the emission we see at all positions in
NGC~7023. Indeed, if the emission at 19$\mu$m is due to C$_{60}$,
there should also be associated features at 7.0, 8.6 and
17.4$\mu$m. Unfortunately, there is severe blending with other PAH
features, most notably the strong 8.6$\mu$m and the 17.4$\mu$m PAH
bands. However, from inspecting the SL spectra, we can clearly see the
7.0$\mu$m C$_{60}$ feature in the wing of the 7.7$\mu$m PAH band at
position 1, and weakly also in position 2; not in the other
positions. It may thus well be that some of the 19$\mu$m emission we
see is due to dPAHs. Finally, we also studied the SL data around the
5.5$\mu$m region. We did not find evidence for an emission feature at
5.5$\mu$m -- if any, they would be even weaker than the already fairly
weak 5.25 and 5.7$\mu$m PAH bands. This is in line with expectations
though: also in our average spectrum, the 5.5$\mu$m band is a few
times weaker than the 10.6$\mu$m band (see
Fig.~\ref{fig:averagespec}).


\section{Discussion}

Several authors have presented detailed theoretical studies on
hydrogenation and dehydrogenation of interstellar PAHs
\citep{1996A&A...305..616A, Vuong-Foing, LePage:hydrogenation,
 2013A&A...552A..15M}. These studies show that interstellar PAHs are
generally either fully hydrogenated, or fully dehydrogenated;
intermediate hydrogenation states can only exist over a very small
range in physical conditions (typically measured by $n_{\rm
 H}/G_0$). This is a consequence of the fact that more dehydrogenated
PAHs are less photostable; thus, once the first H atom is removed, it
becomes easier to remove the next. The hydrogenation balance
furthermore also depends critically on the size of the PAHs. Indeed,
larger PAHs are less susceptible to H loss than smaller PAHs since
they can more easily dissipate the absorbed UV energy. Thus, given the
physical conditions, there is a critical PAH size: PAHs smaller than
this critical size will be rapidly stripped of their hydrogen atoms;
larger PAHs will remain fully hydrogenated. There may be some
discussion about where precisely this size limit is. For NGC~7023 for
instance, \citet{2012PNAS..109..401B} find a critical size of about 60
carbon atoms in the PDR where PAH emission dominates, and about 70
carbon atoms closer to the star where C$_{60}$ is located. However,
\citet{2013A&A...552A..15M} find that even in the PDR (of NGC~7023),
these PAHs are fully dehydrogenated.

There is thus little doubt that a considerable amount of PAH molecules
becomes fully dehydrogenated in interstellar environments. However,
these dPAHs are then expected to further undergo photodissociation,
fragmentation, and isomerization. Thus, it is not expected that a
large population of dPAHs could exist in the interstellar medium. Also
\citet{2008A&A...489.1183M} concluded that there can only be at most a
small population of fully dehydrogenated species in the ISM since
dPAHs would otherwise leave a clear imprint on the UV extinction
curve.

If the emission at 10.68$\mu$m and 19$\mu$m that we observe in
NGC~7023 is indeed due to dPAHs, it implies that there is a small, but
observable population of fully dehydrogenated PAH molecules in these
environments. This could reflect that dPAHs survive just long enough,
or alternatively could point to a subpopulation of dPAHs that are more
hardy and withstand radiation more easily than others. It is puzzling
though that the strength of the 10.68$\mu$m feature is so constant
from position 2 to position 6 while the PAH bands at the same time
change in intensity by a factor of 10. It appears as if the changing
physical conditions ($n_{\rm H}, G_0$) have little to no effect on the
emitting population until reaching position 1 where the dPAHs
disappear. Most likely this represents the position where even the
hardiest species photodissociate; alternatively, this is where they
isomerize into cages and develop into fullerenes. If this is the case,
the fullerenes likely form from smaller cages (since smaller dPAHs at
this point are more abundant due to fragmentation) rather than from
larger PAHs. Position 1 could then represent the point were the
conditions are energetic enough for close network growth (CNG) and
isomerization into C$_{60}$ to complete.

\section{Summary \& Conclusions}

We have studied the IR spectroscopic properties of fully
dehydrogenated PAH molecules. We have created a complete database of
fully optimized PAH structures for species with up to eight benzenoid
rings, and their fully dehydrogenated counterparts. The majority
(64\%) of the PAHs and about half of the dPAHs in our database turn
out to be non-planar. Upon dehydrogenation, most (71\%) dPAHs develop
pentagons in their carbon skeleton. We then calculated the frequencies
and intensities of the normal modes for all species in our database
and analyzed their spectra. We found that dPAHs characteristically
exhibit features at 5.5$\mu$m, 10.6$\mu$m and 18.9$\mu$m in addition
to a forest of weaker features in the 16--30$\mu$m range that appear
as structured continuum emission. We searched for these features in
spectra of the well-studied reflection nebula NGC~7023, corresponding
to various positions with changing physical conditions and hence
different emission characteristics. We detect weak emission at
10.68$\mu$m in 5 out of 6 positions; this emission is furthermore
accompanied by emission at 19$\mu$m that we see at all positions. We
tentatively see this as evidence for a small population of emitting
dPAHs in these environments. These emission characteristics disappear
closest to the illuminating star, where C$_{60}$ has been
detected. This suggests that C$_{60}$ could indeed form from dPAHs,
but it appears more likely that they form from smaller dPAHs rather
than from larger species. 

\acknowledgements 

We would like to thank Xander Tielens and Lou Allamandola for the
stimulating and helpful discussions. We would also like to thank the anonymous referee for the constructive comments.
This work was supported by a
grant from the Academic Development Fund (ADF) from Western
University, and by a Discovery Grant from the Natural Sciences and
Engineering Research Council (NSERC). This work was made possible by
the facilities of the Shared Hierarchical Academic Research Computing
Network (SHARCNET:www.sharcnet.ca) and Compute/Calcul Canada.


\end{document}